\documentclass[sigconf,nonacm]{acmart}

\usepackage{hyperxmp}
\usepackage{tikz}

\AtBeginDocument{%
  \providecommand\BibTeX{{%
    \normalfont B\kern-0.5em{\scshape i\kern-0.25em b}\kern-0.8em\TeX}}}

\makeatletter
\def\@ACM@checkaffil{% Only warnings
    \if@ACM@instpresent\else
    \ClassWarningNoLine{\@classname}{No institution present for an affiliation}%
    \fi
    \if@ACM@citypresent\else
    \ClassWarningNoLine{\@classname}{No city present for an affiliation}%
    \fi
    \if@ACM@countrypresent\else
        \ClassWarningNoLine{\@classname}{No country present for an affiliation}%
    \fi
}
\makeatother

\begin{document}

\title{S3C2 Summit 2024-08: \\ Government Secure Supply Chain Summit}

\author{Courtney Miller$^{\mathsection}$, William Enck$^{*}$, Yasemin Acar$^{\dagger}$, Michel Cukier$^{\ddagger}$, \\ Alexandros Kapravelos$^{*}$, Christian Kästner$^{\mathsection}$, Dominik Wermke$^{*}$, Laurie Williams$^{*}$}

%% authors that will appear in the ACM reference format (without authormarks)
\def \authors{Courtney Miller, William Enck, Yasemin Acar, Michel Cukier, Alexandros Kapravelos, Christian Kästner, Dominik Wermke, Laurie Williams}

\affiliation{%
    \institution{ $^*$North Carolina State University, Raleigh, NC, USA}
}
\affiliation{%
    \institution{$^\dagger$Paderborn University, Paderborn, Germany and George Washington University, DC, USA}
}
\affiliation{%
    \institution{$^\ddagger$University of Maryland, College Park, MD, USA}
}
\affiliation{%
    \institution{ $^\mathsection$Carnegie Mellon University, Pittsburgh, PA, USA}
}
%\email{{firstname.surname}@uni-paderborn.de}
%\email{mcukier@umd.edu}
%\email{{whenck, akaprav, lawilli3}@ncsu.edu}
%\email{kaestner@cs.cmu.edu}

\renewcommand{\shortauthors}{Secure Software Supply Chain Center (S3C2)}
\renewcommand{\shorttitle}{S3C2 Summit 2023-06: Government Secure Supply Chain Summit}

\begin{abstract}
  
  Software supply chains, while providing economic and software development value, are only as strong as their weakest link. Over the past several years, cyberattacks have increased exponentially, explicitly targeting vulnerable links in critical software supply chains. These attacks disrupt the day-to-day functioning and threaten the security of nearly everyone on the internet, from billion-dollar companies and government agencies to hobbyist open-source developers. 
  The ever-evolving threat of software supply chain attacks has garnered interest from the software industry and the US government in improving software supply chain security. 
  On Thursday, August 29th, 2024, three researchers from the NSF-backed Secure Software Supply Chain Center (S3C2) conducted a Secure Software Supply Chain Summit with a diverse set of 14 practitioners from 10 government agencies. 
  The goals of the Summit were to:
  (1)~to enable sharing between participants from different government agencies regarding practical experiences and challenges with software supply chain security;
  (2)~to help form new collaborations;
  (3)~to share our observations from the two summits conducted with industry in the past year; and 
  (4)~to learn about the participants' challenges to inform our future research directions. 
  The summit consisted of discussions of six topics relevant to the government agencies represented, including software bill of materials (SBOMs) and vulnerability exploitability exchange (VEX); updating vulnerable dependencies; malicious commits; reducing entire classes of vulnerabilities; culture; and large language models (LLMs). 
  For each topic of discussion, we presented a summary of the takeaways from our previous industry summits and a list of questions to participants to spark conversation. 
  In this report, we provide a summary of the summit. 
  The open questions and challenges that remained after each topic are found at the end of each topic's section, and the initial discussion questions for each topic are in the appendix.

\end{abstract}

\iffalse

\fi



%\received{30 September 2022}
%\received[revised]{1 December 2022}

\maketitle

\begin{tikzpicture}[overlay, remember picture]
\node[anchor=north west, %anchor is upper left corner of the graphic
      xshift=17.5cm, %shifting around
      yshift=-2.1cm] 
     at (current page.north west) %left upper corner of the page
     {\includegraphics[width=2.1cm]{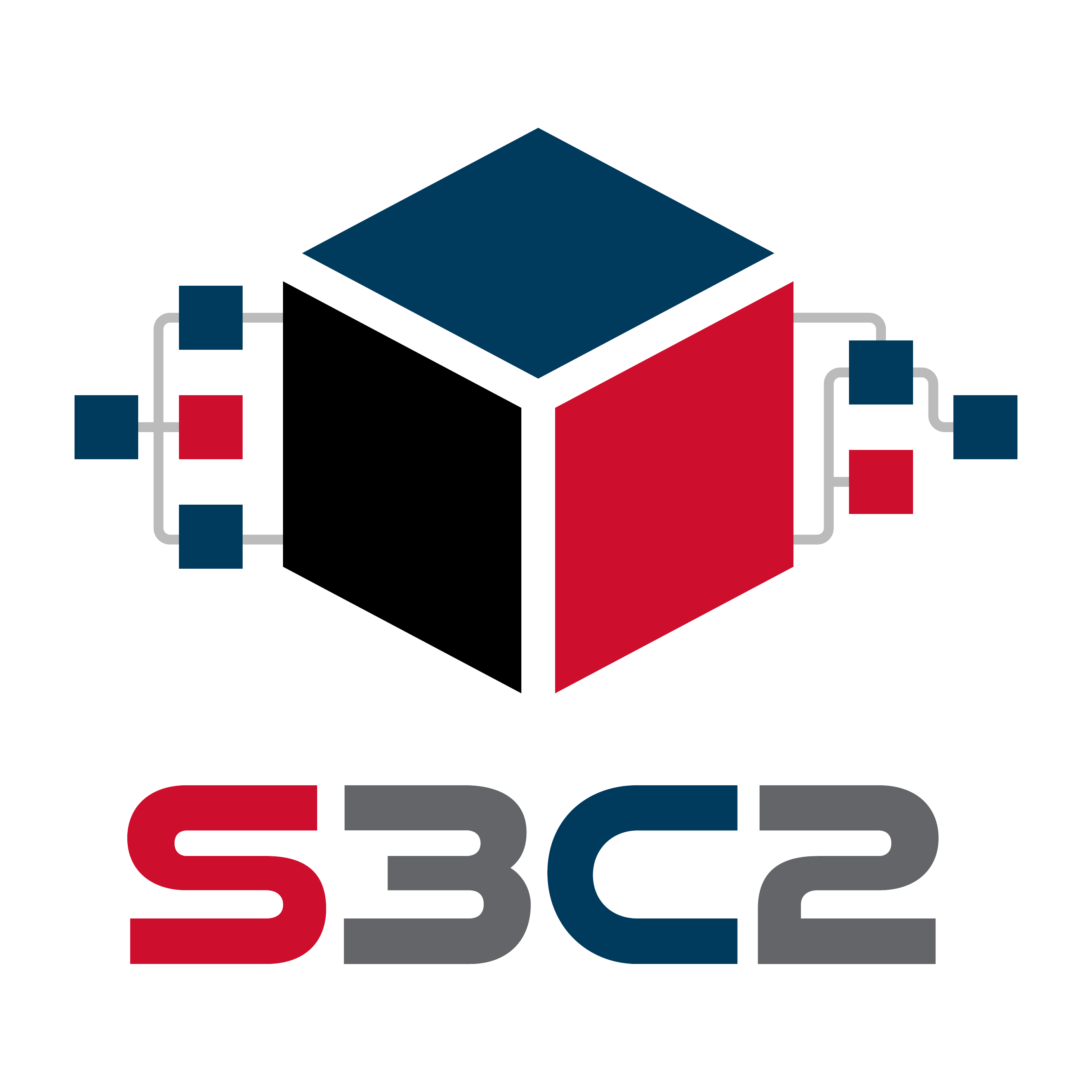}}; 
\end{tikzpicture}

\section{Introduction}

Software supply chains are only as strong as their weakest link. Over the past several years, there has been an exponential increase in cyberattacks specifically targeting vulnerable links in critical software supply chains, disrupting the day-to-day functioning and threatening the security of nearly everyone on the internet, from billion-dollar companies and government agencies to hobbyist open-source developers~\cite{SonatypeReport23}. 
The rapid development of state-of-the-art artificial intelligence (AI) integration systems and large language models (LLMs) has also presented additional novel attack vectors for software supply chain attacks. 
The ever-evolving threat of software supply chain attacks has garnered interest from the software industry and the US government in improving software supply chain security. 

On Thursday, August 29th, 2024, three researchers from the NSF-backed Secure Software Supply Chain Center (S3C2)\footnote{https://s3c2.org/} conducted a day-long Secure Software Supply Chain Summit with a diverse set of 14 participants from 10 government agencies. 
The goals of the Summit were to:
  (1)~to enable sharing between participants from different government agencies regarding practical experiences and challenges with software supply chain security;
  (2)~to help form new collaborations;
  (3)~to share our observations from the two summits held with industry over the past year ~\cite{Summit1,Summit2,Summit3}; 
  and 
  (4)~to learn about the participants' challenges to inform our future research directions. 

The Summit was run under the Chatham House Rule, meaning all participants could freely use the information discussed. However, disclosing who was present, their affiliations, or who said what is forbidden. As such, this report also follows the Chatham House Rule.  
Summit participants were recruited from 10 government agencies invested in software security. 
Attendance was intentionally capped to create an environment that encourages candid conversations among key stakeholders. 

The Summit consisted of discussions of six topics that were decided ahead of time by the participants by voting on which topics to discuss.
The voting process ensured that the topics were of interest and relevant to the government agencies represented. 
The discussion topics included software bill of materials (SBOMs) and vulnerability exploitability exchange (VEX); updating vulnerable dependencies, malicious commits; reducing entire classes of vulnerabilities; culture; and large language models. 
Each topic was moderated by one of the S3C2 researchers, beginning with a brief introduction, a summary of the takeaways from our industry summits, and a list of questions to spark conversation. 
Open questions and challenges for each topic can be found at the end of each section, and all questions posed to practitioners are available in the appendix. 

Three S3C2 researchers (two professors and one PhD student) took notes on the discussion. The PhD student created a first draft of this report based on these notes, which the two professors then reviewed and revised. 
The Summit participants then reviewed the draft. 

The remaining sections of this report summarize this Secure Software Supply Chain Summit.

\section{SBOM and VEX}
\label{sec:sbom}
US Executive Order 14028 requires all organizations and contractors selling software to the federal government to provide a software bill of materials (SBOM) for said software~\cite{EO}. SBOMs are a comprehensive list of all software components and dependencies present in a given software artifact, providing consumers with more transparency about the third-party components they rely on directly and transitively. 
However, not all vulnerabilities are created equal, and just because vulnerabilities are present in a dependency does not mean they affect the software or are actually reachable. A Vulnerability Exploitability eXchange (VEX) is a `companion artifact' and is a form of a security advisory that indicates whether a product is affected by a known vulnerability or vulnerabilities, thus supporting more effective use of SBOMs~\cite{VEX}. 

\subsection{Trust and VEX}

While VEX artifacts help support vulnerability triage and reduce the manual effort required to assess whether every vulnerability that could impact a product does so, there were widespread concerns regarding trust as consumers of VEX artifacts. 
Participants pointed out that it is in the best interest of the large companies leading the charge on the production of VEX artifacts to show that the vulnerabilities in question are not, in fact, an issue for their products. 
For participants, this led to a question of trust. 
This lack of trust is further exacerbated by participants' complaints that such systems currently rely on a lot of trust and very few actual verification steps. One participant expressed interest in employing audits to assess whether the companies producing VEX artifacts are doing what they say they are doing.  
Participants are not currently sure which VEX producers are trustworthy, and several expressed a desire for increased transparency regarding VEX artifacts that have been proven inaccurate. 
So we can collectively learn from VEX artifacts that are proven inaccurate and develop a sense of which VEX producers are actually empirically proven to be trustworthy.

\subsection{Consuming SBOMs and VEX}

Participants reported experiencing some challenges related to producing and consuming SBOMs in classified settings.
For example, consuming SBOMs from a weapons system is difficult.
The SBOM can often not be consumed on the same network where it is produced because if a vulnerability is discovered, it raises the information's classification level.
Furthermore, the information becomes top secret if a vulnerability impacts multiple weapons.
Existing SBOM tools will cause spillage incidents, requiring networks to be shut down.
%, often because of security classification guidelines that make life painful. 
%VEX artifacts, specifically when embedded into SBOMs, can exacerbate these clearance-related issues. For example, in addition to there being situations where practitioners and their subordinates cannot consume an SBOM on the same network it was produced on, if a certain class of vulnerability is identified in a VEX addendum, that vulnerability's existence can up the classification of the information to a level that is higher than that of the original developer working on the system, causing logistical issues.  

A participant noted that newer SBOM standards can have vulnerability information embedded directly into the SBOM.
This led to a discussion on associating vulnerability information with a given SBOM.
Some participants argued for an approach where SBOMs only contain static data about the software component once it is built, and dynamic VEX data (which often changes daily) is stored in an independent document that refers back to the SBOM itself. 
Such an approach would improve clearance-related issues by allowing practitioners to access the original SBOM without interacting with vulnerability information. 
This approach would be similar to Germany's SBOM policies, which prohibit the inclusion of vulnerability information in SBOMs~\cite{germanSBOMPolicyGuideline}.
Whether embedding VEX information directly into SBOMs is beneficial depends on an agency's specific data usage, leaving participants split on this topic.

Several participants also emphasized the need to address foreign nations contributing to open-source digital infrastructure and what that means for usage. For example, one participant pointed out that China significantly contributes to Kubernetes, but does that mean we should care? Are we not going to use Kubernetes?
Participants called for a transition from the presence of foreign nations in open source supply chains being a boolean flag to a quantifiable risk.

Some participants also echoed industry concerns about publicly available SBOMs potentially increasing security risks. One participant described how, while a single unclassified SBOM is not a useful target for a foreign nation, a huge repository of all SBOMs for civilian systems could be a valuable target.

\subsection{Progress Toward Self-Attestation}

As one participant put it, the concept of self-attestation is sound, the execution, not so much. 
Participants report agencies experiencing challenges with self-attestation because the SSDF is a framework that gives guidelines but not concrete requirements, and there is no discrete checklist to follow~\cite{SSDF}. 
There are also modifications and hoops that agencies have to jump through. 
Some agencies are collecting data but are not willing to upload that data to the central repository due to the additional work required.
Furthermore, when industry contractors have to meet the self-attestation requirements for their vendors, it can be difficult to get them to reach full compliance—with some pushing back and requesting additional compensation for meeting self-attestation requirements. 
A participant noted that some companies are just making their projects open source so that they do not need to do the attestation.
This was considered a good thing, as we all get to look at the source code.

Participants also questioned enforcement. Currently, some software suppliers are taking a reactive stance toward self-attestation. 
There were questions about which government agency has the resources to enforce these policies, like the IRS does for taxes.

\subsection{Open Questions}
% \begin{itemize}
% \item From your perspective, how widespread is the practice of: Producing SBOMs? Consuming SBOMs? Storing SBOMs? Sharing SBOMs?
% \item What challenges are you facing? 
% \item What will/can SBOMs actually achieve? How can they be leveraged/used?  
% \item Are you seeing VEX being used?
% \item Do you see VEX being helpful or hurtful? In what ways?
% \item How is the implementation self-attestation going?
% \end{itemize}

\begin{itemize}
\item How can the trustworthiness of VEX artifacts and producers be evaluated, verified, and made more transparent to consumers? 
\item What are best practices for managing SBOM documents with vulnerability information over their life-cycle, particularly in classified environments, that significantly complicate their production and consumption?
\item How can we effectively communicate vulnerability risks in dependencies to developers when the existence of a vulnerability is a higher security clearance than the developer?
\item How can we quantify the risk associated with foreign nation contributions in software supply chains and transition from contributions from foreign nations being a boolean flag?
\item How can self-attestation requirements be enforced across the government? Which government agency has the resources to perform audits and enforce these policies?  
\end{itemize}

\section{Updating Vulnerable Dependencies}
\label{sec:update}

Modern software relies on dependencies as building blocks, allowing for rapid reuse and lower upfront development costs. However, dependencies also have drawbacks, namely dependency management. Keeping up with dependency vulnerability patches can be overwhelming and requires significant manual effort from already overburdened developers. It can be difficult to determine which vulnerabilities are necessary to invest time into addressing and which are not, leading to what some refer to as \textit{patch fatigue}. 

\subsection{Current State of the Practice}

Some participants report having to go through complex software review board processes that can sometimes take six to nine months to get approval for a certified dependency update. 
This leaves some participants stuck in a bureaucratic process that significantly hinders the technical updating process. 
Sometimes, it is more efficient to write code from scratch in-house instead. 
Several participants argued that moving forward, program managers must incorporate supply chain security considerations into their risk based decisions, the same way they consider cost, schedules, and performance.

Similar to the industry, participants reported observing increased use of software component analysis (SCA) tools, which can aid in the identification and management of dependency vulnerabilities. They are also receiving an increasing number of requests for SCA tools, particularly from newer program offices. 
Some participants had positive things to say about SCA tools, with a feature that one practitioner especially appreciated: visualizing the delta code change in a dependency's new release. 
However some participants cautioned that an over-dependence on such tools by software quality assurance personnel without a formal education in computer science may lead to undesired outcomes. 
Furthermore, quality assurance personnel drive the adoption of such tools to provide feedback to developers. Still, the analysis and any vulnerability information identified as a part of it can increase the classification of the information to a level higher than that of the developers on the product team, making it difficult to actually respond to and address vulnerable dependencies. 
Similarly, when vulnerabilities impact multiple agencies, it can be difficult logistically to manage triage because government agencies are siloed. 

Generally speaking, the majority of the challenges described by the participants in this discussion were not technical but rather organizational and bureaucratic.

\subsection{Open Questions}

\begin{enumerate}
    \item How can the logistical management of vulnerability triage that affects multiple government agencies be done more efficiently and collaboratively?  
    \item What practices can reduce bureaucratic overhead in the technical process of updating software library dependencies?
    \item How can agencies benefit from SCA and similar tools without their reports creating unnecessary bureaucratic burdens on software developers?
    \item How can SCA report information be shared with developers while simultaneously accounting for clearance levels and siloed agencies?
\end{enumerate}

\section{Malicious Commits}
\label{sec:malicious}

Instead of waiting for the identification of an existing vulnerability to exploit, attackers are increasingly using malicious commits in software supply chains as an attack vector. 
Through the contribution of malicious commits to a project, attackers can create vulnerabilities themselves and then exploit them. 
An example of this is the recent incident in March of 2024 involving XZ Utils, a file compression library used by Linux distributions in systems worldwide, including Red Hat and Debian. 
In the XZ attack, a malicious actor slowly established themselves as an innocuous maintainer of the XZ project and then used their privileges to gradually create a backdoor, allowing attackers unauthorized access to systems running compromised versions~\cite{xzMediumArticle}. 
The backdoor was accidentally identified before it became widely deployed, but the incident still caused global shock waves and highlighted the need to improve strategies for detecting and mitigating advanced persistent threats (APTs).

\subsection{The Current State of Practice}

While the malicious commit in the XZ incident was caught before deployment, participants pointed out that there were almost certainly countless more similar attacks that were not, leading to questions about how to proceed in such a landscape. 
One participant cited the practices of an agency that, rather than using the traditional model of viewing their system as a walled garden with firewalls and security measures to keep adversaries out, works under the assumption that adversaries have already infiltrated their system. Instead, they focus on design strategies to ensure the continued normal functioning of their system while adversaries are `mucking about.'

Participants pointed out that an underlying issue related to these attacks is that usually when you incorporate libraries into a product, they have all the product's privileges.
Hence, such libraries are used as attack vectors for malicious commits.
%are given unnecessary privileges. 
For example, why did a commit in XZ, a file compression library, have the privilege to exploit SSH in the first place? 
Some participants suggested using strategies such as component isolation and sandboxing to address the technical issue presented by malicious commits, noting that they are not a solution to malicious commits altogether.

Identifying the malicious intent of a commit is challenging.
A participant noted that we need to make it harder to have malicious commits look like errors.
For example, given that there are already many C/C++ vulnerabilities, attackers will make their malicious code look like memory vulnerabilities that can be exploited.
This is a benefit of memory-safe languages.
Another participant noted that malware usually has a payload, which will always be a unique contribution to the malware.
This can help identify intent.

Participants point out that while there is no silver bullet for malicious commits, the integration of strategies to make them harder to perform and easier to detect could lead to meaningful improvements. 
Some detection strategies suggested by participants include code review, scanning, component behavioral analysis, and measuring CPU usage (which is how the XZ backdoor was identified). However, it should be noted that the detection strategy that will be effective in a given strategy will depend on the context. 

\subsection{Open Questions}

\begin{enumerate}
    \item What are best practices for ensuring continued normal system functioning in a landscape where many other XZ-like attacks likely remain identified?
    \item What combination of strategies, including restricting library privileges, component isolation, sandboxing, and memory-safe language usage, could lead to meaningful improvements in preventing malicious commit attacks? 
    \item Given the performance and development overheads of component isolation and sandboxing, how can they most effectively mitigate malicious commit risks?
    \item Given a specific context, how can a determination be made about the best malicious commit detection strategies?
\end{enumerate}

\section{Reducing Entire Classes of Vulnerabilities}
\label{sec:reduce}

Adopting particular types of programming languages or frameworks can reduce a system's risk for entire classes of vulnerabilities. For example, an industry practitioner at a previous summit pointed out that many vulnerabilities are memory-related, so moving to memory-safe languages like Rust can significantly reduce a system's risk for memory-related vulnerabilities. However, doing so can require significant overhead and be challenging to sell to senior leadership as a worthwhile investment.  

\subsection{The Current State of Practice}

Participants report increasing use of secure frameworks, but usually in contexts where systems must perform perfectly reliably. 
New hires recently out of school are also incorporating more secure frameworks where possible, but many of the legacy codes they interact with use lower-level primitives. 

Some participants expressed interest in moving to memory-safe languages like Rust but cited the lack of reliable translation tools as a significant hindrance since translating the code manually is often too costly. 
Furthermore, Rust's steep learning curve is also a concern because it makes recruiting and training developers more difficult. 

Although participants appreciated the benefits of adopting memory-safe languages like Rust, some inherent downsides to Rust, such as the performance trade-offs, made adoption less appealing. 
Additionally, some participants in the Space industry are hesitant to adopt Rust because it tends to throw more errors and faults in high-radiation environments than other languages already in their legacy code, like C and C++.

\subsection{Open Questions}

\begin{enumerate}
    \item How can legacy systems be translated to memory-safe languages like Rust, given the lack of reliable translation tools and the often prohibitively high cost of manual code translation? 
    \item How can the security benefits of translating legacy code systems to memory-safe languages like Rust be balanced with the performance trade-offs and overhead costs?
    \item How can we teach young developers to use secure frameworks while ensuring they understand the fundamentals they need to secure legacy code?
    \item How effective and automated will Rust translation tools become? When would be the right time to adopt them?
\end{enumerate}

\section{Culture}
\label{sec:culture}
A healthy and robust security culture includes shared responsibility across all levels of an organization. There is not just a `security team' with `security people'; rather, everyone is involved in security in the context relevant to them. 
Recent legislation has changed the workflow of software development organizations, but it is unclear if the culture has caught up.

\subsection{Organizational Security Culture}
Participants report observing a cultural change that promotes security as an enabler rather than a roadblock to the development of high-quality software. 
Some participants have had success conveying this cultural shift to personnel in their agencies when they drop the security vernacular and instead communicate both the security needs and the consequences of not meeting those needs using language, analogies, and examples that resonate with each personnel member —connecting supply chain monitoring to vulnerability risk. 
Making the security needs more personal to each personnel supports their internalization of how the organization's security needs directly relate to and impact the things they care about. 
However, even with effective communication, building the culture will take time, and progress will be slow.

\subsection{Software Security as a Liability}
Some participants called for a change in how we think about risk and liability related to cybersecurity events in the United States, pointing to Europe's model as an example. 
In Europe, software is considered a product, which allows the government to mandate the liability for cybersecurity events to the companies that produced the software, making companies more culpable for cybersecurity events that occur in their products to safeguard consumers~\cite{EUCyberResilienceAct}.
However, in the United States, software is largely unregulated, making the delegation of liability for cybersecurity events much more unclear, with litigation currently being the primary method of settling such disputes.

Along these lines, several participants also called for standardizing security engineering practices. 
They point out that, as United States software regulations currently stand, no laws place liability on software producers when and if their products fail in completely benign standard environments, much less in any other context. 
However, some participants pushed back on this, arguing that if such legal liabilities were put in place, very few people would build software because of the risk involved. 
Participants agreed that observing how the rollout of the European Union's recent Cyber Resilience Act (CRA) could help inform ideas for similar legislation in the United States.

\subsection{Open Questions}

\begin{enumerate}
    \item What lessons will the European Union's rollout of the Cyber Resilience Act provide for legislation in the United States, and how will those lessons impact how we think about risk and liability related to cybersecurity events? 
    \item What are best practices in security engineering? Should they be standardized in the United States?
\end{enumerate}

\section{Large Language Models (LLMs)}
\label{sec:llm}
In recent years, many AI-enabled tools leveraging LLMs have been released onto the market, such as Claude and Copilot, which many developers have quickly embraced. Developers are increasingly using LLMs to generate and analyze existing code. However, we are still in the process of determining which use cases benefit from the use of LLMs and which do not.

\subsection{LLMs as a Development Support Tool}
Participants' perspectives on using LLMs in their agencies varied widely depending on their use case. 
Participants reported several positive use cases for LLMs that they had either experienced themselves or heard about from another agency, including generating assurance proofs, creating personas for attacks, detecting malicious commits, fixing documentation problems, and fixing code readability. 
However, because the risks of LLM use are poorly understood and highly imminent, there was a desire to keep LLMs far away from more sensitive artifacts like weapons systems.

\subsection{LLMs as an Attack Vector}
Some participants expressed concerns about the inevitability of LLMs being leveraged as future software supply chain attack vectors. 
For example, LLMs could generate a large volume of innocuous commits as noise to hide a real threat under, essentially DOSing software maintainers with an avalanche of pull requests.

\subsection{Open Questions}

\begin{enumerate}
    \item How can risks associated with LLMs be clearly communicated with developers so that they can appropriately leverage them in their environments?
    \item Can defenses be proactively deployed to mitigate expected LLM-based supply chain threats, such as DoSing software maintainers with pull requests?
\end{enumerate}

\section{Executive Summary}
As software consumers, some agencies experienced challenges consuming SBOMs and VEX artifacts, including clearance-related issues. 
Progress toward self-attestation has been difficult due to a lack of concrete requirements, and some have struggled to get industry contractors to fully comply with self-attestation vendor requirements.
Similar to industry, there has been an increase in the use of SCA tools to support the detection and mitigation of vulnerabilities in dependencies; however, the most significant hindrances to these processes were not technical but rather organizational and bureaucratic. 
The XZ incident highlighted the need to change the way practitioners reason about system security. While there is no silver bullet for malicious commits, integrating strategies to make them harder to perform and easier to detect could lead to meaningful improvements. 
While the use of secure frameworks and memory-safe languages, such as Rust, can reduce entire classes of vulnerabilities, logistical issues often hinder their adoption at a wide-scale. 
Participants have observed a cultural shift where security is being seen as an enabler rather than a roadblock, but cultural change takes time and progress is slow and steady.
LLMs are viewed as both a useful tool in certain contexts and a potential vector for supply chain security attacks.

\section{Acknowledgements}
A big thank you to all Summit participants. We are very grateful for hearing about your valuable experiences and suggestions. Laurie Williams and William Enck organized the summit, which was recorded by Courtney Miller. This material is based upon work supported by the National Science Foundation Grant Nos. 2207008, 2206859, 2206865, and 2206921.
These grants support the Secure Software Supply Chain Summit (S3C2), which consists of researchers at North Carolina State University, Carnegie Mellon University, University of Maryland, and George Washington University. Any opinions expressed in this material are those of the author(s) and do not necessarily reflect the views of the National Science Foundation.
%\end{acks}

\bibliographystyle{ACM-Reference-Format}
\bibliography{literature}

\appendix

\section{Initial Discussion Questions}
\label{questions}
\begin{enumerate}
\item \textbf{Software Bill of Materials} From your perspective, how wide-spread is the practice of: producing SBOMs? Consuming SBOMs? Storing SBOMs? Sharing SBOMs? What challenges are you facing? What will/can SBOMs actually achieve? How can they be leveraged/used? Are you seeing VEX being used? Do you see VEX being helpful or hurtful?  In what ways? How is the implementation of self-attestation going? 

\item \textbf{Vulnerable dependencies} From the perspective of the government agencies you interact with, what are their main concerns and pain points around updating vulnerable dependencies? Do they have policies around when to update? What kind of testing or other strategies are used before updating to a new version?  How do SCA tools fit into the decision?

\item \textbf{Malicious commits} How seriously are the government agencies you interact with consider malicious commits? How can malicious commits be detected? What do you think signals a suspicious/malicious commit?  What role does the ecosystem play in detecting malicious commits?

\item \textbf{Reducing entire classes of vulnerabilities} Are you moving toward the use of safer languages?  Mandating the use of any secure frameworks?

\item\textbf{Culture} What changes have you made to support supply chain security/executive order compliance?  What do you think is needed for nurturing such a security-benefiting culture?
 
\item \textbf{LLMs and Supply Chain Security} From the perspective of government agencies you interact with, what is your perspective on the use of large language models (LLMs) such as ChatGPT as another supply chain attack vector?
\end{enumerate}

\end{document}